\documentstyle[12pt]{article}

\begin{document}
{\sf \begin{center} \noindent
{\Large \bf Lagrange Instability of Geodesics in Curved Double Twisted Liquid Crystals}\\[3mm]

by \\[0.3cm]

{\sl L.C. Garcia de Andrade}\\

\vspace{0.5cm} Departamento de F\'{\i}sica
Te\'orica -- IF -- Universidade do Estado do Rio de Janeiro-UERJ\\[-3mm]
Rua S\~ao Francisco Xavier, 524\\[-3mm]
Cep 20550-003, Maracan\~a, Rio de Janeiro, RJ, Brasil\\[-3mm]
Electronic mail address: garcia@dft.if.uerj.br\\[-3mm]
\vspace{2cm} {\bf Abstract}
\end{center}
\paragraph*{}
It is shown that curved and flat helical double twisted liquid
crystal (DTLC) in blue phase, can be unstable (stable) depending of
the sign, negative (positive) of sectional curvature, depending on
the pitch of the helix of the nematic crystal. In both cases Cartan
torsion is presented. It is also shown that the instability or
stability depends on the value of the pitch of the helix in nematic
crystals. Frank energy stability A similar result using the method
of Frank energy stability in the twist of cholesteric liquid crystal
was given by Kiselev and Sluckin [PRE 71(2005)], where the free
twist number determines the equilibrium value of the cholesteric
liquid crystals (CLC) pitch of the helix. As a final example we
solve the geodesic equations in twisted nematics with variable pitch
helix and non-constant torsion. This non-Riemannian geometrical
approach, also seems to unify two recent analysis of cylindrical
columns given by Santangelo et al [PRL 99,(2007)] and the curved
crystal endowed with torsion, given by Vitelli et al [Proc Nat Acad
Sci (2006)]. Stability of toroidal curved surfaces were also
previously considered by Bowick et al [PRE 69,(2004)], as an example
of curvature-induced defect. Investigation of the Lagrangean
instability may be useful in the investigation of HIV viruses and
proteins.\vspace{0.5cm} \noindent {\bf PACS numbers:}
\hfill\parbox[t]{13.5cm}{02.40.Hw:Riemannian geometries}

\newpage
\newpage
 \section{Introduction}
 Following the steps of success of the applications of the
 differential geometry of curves and surfaces \cite{1}, through Riemannian geometry in Einstein
 theory of gravitation \cite{2}, as well as the success of the use of Cartan
 non-Riemannian geometry endowed with a torsion tensor, to explain
 alternatives theory of gravitation such as Einstein-Cartan gravity \cite{3},
 many mathematical and crystal physicists, have applied these
 geometries to the explanation of disclination (Riemann curvature)
 or dislocation (non-Riemannian Cartan torsion), defects in crystals
 \cite{4}. Even recently Riemannian geometry has been applied in the
 investigation of twisted magnetic flux tubes in plasmas helical
 flows MHD with and without vortices \cite{5,6,7}. More recently these paths, two recent papers on the stability of geodesics in the curved and flat
 cylindrical liquid crystals in curved surface with defects \cite{8} and columnar phases \cite{9} have
 appeared in the literature. In the first paper Vitelli et al \cite{8} have
 discussed the instability driven by a purely geometric potential in Gaussian curvature of the stress law covariant
 generalization. In the second Santangelo et al \cite{9} have been considered the instability or stability of geodesics in
 the geometry of columnar phases of nematic liquid crystal, in the analogy with the focusing process of a lensing ray in optics
 or gravitational lensing. Recently we also use the sectional developed by Kambe \cite{10}
 curvature method to investigate the Lagrangean instability of a Couette type flow, viscous and sheared.
 Actually Kambe \cite{10} has previously investigate Couette planar flows stability, and showed that Riemann curvature tensor
 would exist in the case of existence of pressure, even a constant one. The stability of the Couette planar flow is also
 obtained by him, using the symmetry in the case of symmetric scalar stress potential and by making use of the technique
 of Riemann sectional curvature, where the negative sectional curvature indicates instability of the flow, in the
 Lagrangean sense. In this brief report this idea can be easily transported to investigate the instability of DTLC nematic blue
 phase, where as shown by Pansu and Dubois-Violette \cite{11} the presence of torsion is taken from granted from the very
 beginning. As a last example we modify the constant torsion twisted nematic geometry with variable pitch and compute the
 geodesics for this CLC. Therefore we investigate the Lagrange instability in two simple examples, in analogy with the works of Vitelli
 et al and Santangelo et al, namely of a Riemann-flat torsioned  DTLC , while
 in the second example a curved crystal is considered. The usefulness
 of the investigation of the Riemannian and non-Riemannian geometries
 in physical allows one also building models for analog gravity as
 the case of non-Riemannian geometry of vortex acoustic
 flows \cite{12}. Other curvature-induced defect in toroidal Riemann
 geometry were also investigated by Bowick et al \cite{13}. Yet another example of the use of Gaussian curvature on continuum
 elastic model topological defect in Riemannian manifolds was investigated by Giomi and Bowick \cite{14}. The paper is organized as follows: Section II presents a brief
 review of sectional Riemannian curvature method in the
 coordinate-free language which helps to unify previously examples of instability.
 Section III presents the examples of curved and flat
 DTLC with totally skew torsion Lagrange stability and the
 dependence of the instability on the value of the pitch of helix, a similar result obtained by Kiselev and
 Sluckin \cite{15}. Section IV presents the computation of geodesics of twisted nematic geometry. In section V
 conclusions are presented.
 \section{Sectional Riemannian curvature Instability}
 In this section, before we add we make a brief review of the differential geometry of surfaces in coordinate-free language.
 The Riemann curvature is defined by
 \begin{equation}
 R(X,Y)Z:={\nabla}_{X}{\nabla}_{Y}Z-{\nabla}_{Y}{\nabla}_{X}Z-{\nabla}_{[X,Y]}Z\label{1}
 \end{equation}
where $X {\epsilon} T\cal{M}$ is the vector representation which is
defined on the tangent space $T\cal{M}$ to the manifold $\cal{M}$.
Here ${\nabla}_{X}Y$ represents the covariant derivative given by
\begin{equation}
{\nabla}_{X}{Y}= (X.{\nabla})Y\label{2}
 \end{equation}
which for the physicists is intuitive, since we are saying that we
are performing derivative along the X direction. The expression
$[X,Y]$ represents the commutator, which on a vector basis frame
${\vec{e}}_{l}$ in this tangent sub-manifold defined by
\begin{equation}
X= X_{k}{\vec{e}}_{k}\label{3}
\end{equation}
or in the dual basis ${{\partial}_{k}}$
\begin{equation}
X= X^{k}{\partial}_{k}\label{4}
\end{equation}
can be expressed as
\begin{equation}
[X,Y]= (X,Y)^{k}{\partial}_{k}\label{5}
\end{equation}
In this same coordinate basis now we are able to write the curvature
expression (\ref{1}) as
\begin{equation}
R(X,Y)Z:=[{R^{l}}_{jkp}Z^{j}X^{k}Y^{p}]{\partial}_{l}\label{6}
\end{equation}
where the Einstein summation convention of tensor calculus is used.
The expression $R(X,Y)Y$ which we shall compute bellow is called
Ricci curvature. The sectional curvature which is very useful in
future computations is defined by
\begin{equation}
K^{Riem}(X,Y):=\frac{<R(X,Y)Y,X>}{S(X,Y)}\label{7}
\end{equation}
where $S(X,Y)$ is defined by
\begin{equation}
{S(X,Y)}:= ||X||^{2}||Y||^{2}-<X,Y>^{2}\label{8}
\end{equation}
where the symbol $<,>$ implies internal product. In the
non-Riemannian (NR) case, the torsion two-form $T(X,Y)$ is defined
by
\begin{equation}
T(X,Y):=
\frac{1}{2}[\bar{{\nabla}}_{X}Y-\bar{{\nabla}}_{Y}X-[X,Y]]\label{9}
\end{equation}
where $\bar{\nabla}$ is the non-Riemannian connection \cite{7}
endowed with torsion. As in EC theory \cite{5} the geodesic equation
does not depend on torsion; only Jacobi deviation equation depends
on torsion which is enough for investigate the role of torsion on
stability. Since the Jacobi equation is given by
\begin{equation}
\frac{d^{2}J}{ds^{2}}=[||{\nabla}_{\vec{t}}\vec{e}_{J}||^{2}-K^{NR}(t,\vec{e}_{J})||J||\label{10}
\end{equation}
where $||\vec{e}_{J}||=1$ and J is the Jacobi field, representing
the separation between geodesics,while $\vec{t}$ is the geodesic
tangent vector. Here $K^{NR}(X,Y)$ is given by
\begin{equation}
K^{NR}(X,Y)= K^{Riem}(X,Y)+2<T(X,Y),\bar{\nabla}_{Y}X>\label{11}
\end{equation}
Here as we shall see bellow the geodesic equation is
${\nabla}_{Y}Y=0$ simplified this expression. Note from this
expression that the instability, or separation of the geodesics in
the flow
\begin{equation}
\frac{d^{2}J}{ds^{2}}\ge{0}\label{12}
\end{equation}
implies that $K^{Riem}<0$ which is the condition for Lagrange
instabillity.
\section{Stability of Curved and flat Liquid Crystal with torsion} In this
section let us address the examples of Riemann-flat TDLC and the
curved LC. In the first case we consider the CLC, where twist of
molecules exists in just one direction, which a ordered phase is
given by the director field
\begin{equation}
n^{x}=0 \label{13}
\end{equation}
\begin{equation}
n^{y}=cospx\label{14}
\end{equation}
\begin{equation}
n^{z}=sinpx\label{15}
\end{equation}
where $q:= \frac{2\pi}{p}$ is the wave number and p is the pitch of
the helix. In the blue phase local constraint of minimum energy is
double twist , which physically means that twist occurs in more than
one direction in space. In the Euclidean frame this equation is
\begin{equation}
{\delta}_{k}n^{j}+p{\epsilon}_{kip}n^{i}{\delta}^{jp}=0 \label{16}
\end{equation}
In the Riemann-flat manifold,  where the Riemann tensor $R(X,Y)Z$
vanishes, or in components
\begin{equation}
{R^{i}}_{jkl}:=0 \label{17}
\end{equation}
In the frame where the covariant derivative of the basis
$({\vec{e}}_{i})$ vanishes or
${\nabla}_{{\vec{e}}_{i}}{\vec{e}}_{j}=0$, where from expression
(\ref{9}) the Cartan torsion reads
\begin{equation}
T({\vec{e}}_{i},{\vec{e}}_{j})=
-[{\vec{e}}_{i},{\vec{e}}_{j}]\label{18}
\end{equation}
The double twist Levi-Civita connection is
\begin{equation}
{\nabla}_{k}{\vec{n}}^{j}={\partial}_{k}{\vec{n}}^{j}+{{\Gamma}^{j}}_{ik}\vec{n}^{i}=0
 \label{19}
\end{equation}
Now computing the general Riemann-Cartan curvature tensor
\begin{equation}
R^{RC}({\vec{e}}_{i},{\vec{e}}_{j})=
{\nabla}_{\vec{e}_{i}}{\nabla}_{\vec{e}_{j}}-{\nabla}_{\vec{e}_{i}}{\nabla}_{\vec{e}_{i}}-{\nabla}_{[{\vec{e}}_{i},{\vec{e}}_{j}]}
\label{20}
\end{equation}
Using the expression for the Cartan torsion 2-form $T(X,Y)$ in terms
of the Lie bracket above, the RC curvature reduces to
\begin{equation}
R^{RC}({\vec{e}}_{i},{\vec{e}}_{j})\vec{e}_{j}=-{\nabla}_{[{\vec{e}}_{i},{\vec{e}}_{j}]}\vec{e}_{j}=
({[{\vec{e}}_{i},{\vec{e}}_{j}]}.{\nabla})\vec{e}_{j}
\label{21}
\end{equation}
therefore to compute this expression one computes the Lie bracket in
terms of Cartan torsion as
\begin{equation}
T({\vec{e}}_{i},{\vec{e}}_{j})=
({\vec{e}_{i}}.{\nabla}){\vec{e}_{j}}-({\vec{e}_{j}}.{\nabla}){\vec{e}_{i}}={[{\vec{e}}_{i},{\vec{e}}_{j}]}={\partial}_{i}\vec{e}_{j}-{\partial}_{j}\vec{e}_{i}=
{T^{k}}_{ij}\vec{e}_{k}
\label{22}
\end{equation}
substitution of these components of torsion into expression
(\ref{21}) yields
\begin{equation}
R^{RC}({\vec{e}}_{i},{\vec{e}}_{j})\vec{e}^{j}=({[{\vec{e}}_{i},{\vec{e}}_{j}]}.{\nabla})\vec{e}^{j}
=({T^{k}}_{ij}\vec{e}_{k}.{\nabla})\vec{e}^{j}=({T^{k}}_{ij}{\partial}_{k}\vec{e}^{j})={T^{k}}_{ij}{T}^{lij}\vec{e}_{l}
\label{23}
\end{equation}
Thus the cholesteric blue phase of liquid crystals possesses the
sectional curvature
\begin{equation}
K(\vec{e}_{i},\vec{e}_{j})=<R(\vec{e}_{i},\vec{e}_{j})\vec{e}_{j},\vec{e}_{i}>=
T^{kij}{T}_{kij}=-p{\epsilon}_{ijk}{\epsilon}^{ijk}=-6p^{2}<0
\label{24}
\end{equation}
since this curvature is negative, the instability of the CLC is
proved in this case. In a more general Riemannian manifold we shall
prove now that depending on the value of the pitch the CLC can be
stable in the Lagrangean sense. This can be done as follows. First
we shall compute the Riemann tensor with the condition of geodesic
${\nabla}_{Y}Y=0$ is given, which simplifies the curvature
expression to
\begin{equation}
R(X,Y)Y=-{\nabla}_{Y}[\bar{{\nabla}_{X}}Y+pY{\wedge}X]+
p(X{\wedge}Y.{\nabla})Y\label{25}
\end{equation}
where now, we have used the covariant derivative
\begin{equation}
{\nabla}_{X}Y:=\bar{{\nabla}_{X}}Y+pY{\wedge}X]\label{26}
\end{equation}
To simplifies computations we assume that the term
$\bar{{\nabla}_{X}}Y$ vanishes which after some algebra leads to
expression
\begin{equation}
R(X,Y)Y=[-p^{2}+p]Y{\wedge}(X{\wedge}Y)
\label{27}
\end{equation}
Finally from this expression we can write the expression for the
sectional curvature as
\begin{equation}
K(X,Y)=<R(X,Y)Y,X>=[-p^{2}+p]<Y{\wedge}(X{\wedge}Y),X>=[-p^{2}+p]{\alpha}||X||^{2}\label{28}
\end{equation}
Thus this time the sectional derivative can be positive or the
twisted nematic crystal can be geodesically stable, depending on the
value of the helix pitch.
\section{Geodesics of twisted nematic with variable pitch helix}
Let us built the Riemann metric in analogy to Katanaev and Volovich
\cite{16} as the deformation of the crystals $e_{ij}=n_{(j,i)}$
where this deformation represents the perturbation on the metric
$g_{ij}$ given by
\begin{equation}
g_{ij}={\delta}_{ij}+e_{ij} \label{29}
\end{equation}
where ${\delta}_{ij}$ is the delta Kronecker symbol and the vector
$n_{i}$ where ${(i=1,2,3)}$ is the director field of nematic liquid
crystal. we consider here that Riemann curvature tensor vanishes as
in Einstein theory of absolute parallelism, which implies that the
anti-symmetric connection (torsion tensor) is given in terms of the
metric by
\begin{equation}
{T}_{ijk}=g_{jk,i}-g_{ji,k} \label{30}
\end{equation}
Here the comma denotes partial derivatives with respect to the lower
index. Substitution of expression (\ref{29}) into (\ref{30}) yields
the following torsion vector
\begin{equation}
T_{k}=[{\nabla}^{2}n_{k}-{\delta}_{k}(div n)] \label{31}
\end{equation}
therefore and the Weitzenb\"{o}ck condition for teleparallelism on
the curvature Riemann tensor $R_{ijkl}({\Gamma})=0$ ,where
${\Gamma}$ is the Riemann-Cartan affine connection.Later on we shall
make an application of this formula to a special case of nematic
crystal.In the meantime let us compute the geodesics equations for
the corresponding metric of the liquid crystal.The geodesics
equations
\begin{equation}
\ddot{{x}^{i}}+{\Gamma}^{i}_{jk}{\dot{x}}^{j}{\dot{x}}^{k}=0
\label{32}
\end{equation}
and
\begin{equation}
\ddot{{x}^{i}}+{\delta}^{ik}n_{k,lj}{\dot{x}}^{l}{\dot{x}}^{j}=0
\label{33}
\end{equation}
where
${\Gamma}^{i}_{jk}=\frac{1}{2}{\delta}^{il}[g_{lj,k}+g_{lk,j}-g_{jk,l}]$
and we use the Euclidean 3D metric ${\delta}_{ij}$ to raise and
lower indices.Let us now apply these ideas to the pure twist
geometry of the nematic liquid crystals,where the director field is
now given by the components
\begin{equation}
n_{z}=cos{\theta}(y) \label{34}
\end{equation}
and
\begin{equation}
n_{x}=sin{\theta}(y) \label{35}
\end{equation}
Where ${\theta}$ is the twist angle and the planar crystal is
orthogonal to the y-coordinate direction. Substituting expressions
(\ref{33}) and (\ref{34}) into (\ref{35}) one obtains the following
components of the torsion vector $T^{i}_{ki}=T_{k}$
\begin{equation}
T_{y}({\theta})=-{{\partial}_{y}}(div n) \label{36}
\end{equation}
and
\begin{equation}
T_{z}({\theta})=-{\nabla}^{2}n_{z} \label{37}
\end{equation}
Since $div n=n_{x,x}+n_{y,y}+n_{z,z}=0$ , there is no torsion vector
component along the orthogonal direction. Thus the only
non-vanishing torsion component of torsion reads
\begin{equation}
T_{z}({\theta})=-[cos{\theta}({\frac{d{\theta}}{dy}})^{2}+sin{\theta}\frac{d^{2}{\theta}}{{dy}^{2}}]
\label{38}
\end{equation}
Since local equilibrium conditions on the Liquid crystals yields
\cite{11}
\begin{equation}
{\frac{d{\theta}}{dy}}=constant=K \label{39}
\end{equation}
Thus equation (\ref{38}) reduces to
\begin{equation}
T_{z}({\theta})=-K^{2}cos{\theta} \label{40}
\end{equation}
This last expression tell us that the twisted geometry of the
crystal leads to a 3D helical torsion.Geodesics of the twisted
geometry of the liquid crystal leads to the results
\begin{equation}
\ddot{x}+n^{x}_{,yy}{\dot{y}}^{2}=0 \label{41}
\end{equation}
and
\begin{equation}
\ddot{y}=0 \label{42}
\end{equation}
and finally
\begin{equation}
\ddot{z}+n^{z}_{,yy}{\dot{y}}^{2}=0 \label{43}
\end{equation}
Here the dots means derivation with respect to time coordinate. A
simple algebraic manipulation yields
\begin{equation}
\ddot{x}+{K_{1}}^{2}sin{\theta}(y)=0 \label{44}
\end{equation}
and
\begin{equation}
\ddot{z}+{K_{1}}^{2}cos{\theta}(y)=0 \label{45}
\end{equation}
To simplify matters let us solve these equations in the
approximation of small twist angles ${\theta}<<<0$ where above
equations reduces to
\begin{equation}
\ddot{x}+{K_{1}}^{2}{\theta}=0 \label{46}
\end{equation}
and
\begin{equation}
\ddot{z}+{K_{1}}^{2}=0 \label{47}
\end{equation}
Therefore from expression (\ref{47}) one obtains the following
solution
\begin{equation}
z=-{K_{1}}^{2}t+c \label{48}
\end{equation}
where c is an integration constant.To solve the remaining equations
we need an explicit form of the twist angle with respect to time
,this can be obtained from the integration of the expression
${\frac{d{\theta}}{dy}}=0$ which yields
\begin{equation}
{\theta}={K}y+f \label{49}
\end{equation}
where $f$ is another integration constant.Substitution of
$y=K_{0}t+d$ into this equation yields
\begin{equation}
{\theta}=mt+g \label{50}
\end{equation}
where $f$,$m$ and $g$ are new integration constants. Thus one
obtains
\begin{equation}
\ddot{x}=-({K_{1}}^{2}mt+K_{1}g) \label{51}
\end{equation}
Integration of this expression yields
\begin{equation}
x(t)=-\frac{1}{6}({\alpha}t^{3}+{\beta}t^{2}+{\gamma}t+{\delta})
\label{52}
\end{equation}
substitution of the expression $t=\frac{y}{K_{0}}$ into (\ref{52})
yields
\begin{equation}
x=-\frac{1}{6}({\alpha}^{'}y^{3}+{\beta}^{'}y^{2}+{\gamma}^{'}y+{\delta})
\label{53}
\end{equation}
where all the greek letters represent new integration
constants.Therefore the trajectory of the test particles is
represented by a third order polinomial curve.Trajectories of test
particles in domain walls are in general parabolic curves.After we
finish this letter we hear that Dubois-Violette and Pansu \cite{11}
have considered a similar application of teleparallelism to
cholesteric Blue Phase of liquid crystals.Nevertheless in their
paper Cartan torsion is constant
\section{Conclusions} One of the
most physical question in liquid crystal physics is the
investigation of the stability of flows in the Euclidean manifold
${\cal{E}}^{3}$ using the sectional curvature of geodesic deviation.
In this paper we discuss and present the contribution of Cartan
torsion tensor and its role in curved and Riemann flat liquid
crystals in the blue phase. The Lagrange instabilities in several
CLC cases are computed and it is shown that instability may depend
upon the value of twisting number or the pitch of the helix of
nematics. The models discussed here may also be useful in building
the analog models of stability of geodesics in non-Riemannian
theories of gravity.


\newpage
\begin{thebibliography}{18}
\bibitem{1} Manfredo do Carmo, Differential Geometry of curves and
Surfaces (1992) Birkhause.
\bibitem{2} A. Einstein, The meaning of Relativity, Princeton
(1955).
\bibitem{3} E. Cartan, C.R. Acad. Sciences 174
(Paris) (1922) 593. F. W. Hehl, Yu N. Obukhov, Foundations of
Classical Electrodynamics: Charge, Flux and Metric (2003),
Birkhauser.
\bibitem{4} L.R. Rakotomanana, A Geometric Approach to the Thermomechanics of
continua (2003) Birkhauser.
\bibitem{5} L.C. Garcia de Andrade, Phys Scripta 13 (2006).
\bibitem{6} L.C. Garcia de Andrade, Phys Plasmas 13(2006).
\bibitem{7} L. C. Garcia de Andrade, Curvature and Torsion effects
on carrying currents twisted solar loops, (2006) Phys of Plasmas nov
issue.
\bibitem{8} V. Vitelli, J. Lucks, D.R. Nelson, Proc Nat Acad Sc. 103
(2006) 12323.
\bibitem{9}  C. Santangelo, V. Vitelli, R.D. Kamien, D. R. Nelson,
Phys Rev Lett. 99 (2007) 017801.
\bibitem{10} T. Kambe, Geometrical Theory of Dynamical Systems and
 Fluid Flows (2004) World Scientific.
\bibitem{11} E. Dubois-Violette, E. Pansu, Which Universe is Blue Phase, in Geometry in Condensed Matter Physics (1990)
World Scientific.
\bibitem{12} L.C. Garcia de Andrade,Phys Rev D 70,6
(2004) 064004-1.
\bibitem{13} M. Bowick, D. R. Nelson, A. Travesset, Phys Rev E 69
(2004) 041102.
\bibitem{14} L. Giomi, M. Bowick,Phys Rev. B 76 (2007) 054106.
\bibitem{15} A.D. Kiselev, T.J.Sluckin, Phys Rev. E 71 (2005) 031704.
\bibitem{16} Katanaev, Volovich, Ann Physics (1990).
\end{thebibliography}
\end{document}